\documentclass[12pt]{article}
\usepackage{graphicx,amsmath,float}
\usepackage{units}
\usepackage{multirow}
\usepackage{booktabs}
\usepackage{array}
\usepackage{lineno}
\newlength{\dinwidth}
\newlength{\dinmargin}
\def\lapproxeq{\lower .7ex\hbox{$\;\stackrel{\textstyle                                                    
<}{\sim}\;$}}                                                    
\def\gapproxeq{\lower .7ex\hbox{$\;\stackrel{\textstyle                                                    
>}{\sim}\;$}}                                                    
\def\be{\begin{equation}}                                                    
\def\ee{\end{equation}}                                                    
\def\bea{\begin{eqnarray}}                                                    
\def\eea{\end{eqnarray}}

\def\sh{\hat s}
\def\sh2{{\hat s}^2}

\begin{document}
                                                    
\titlepage                                                    
\begin{flushright}                              
%CERN-TH-???\\                       
IPPP/20/23  \\  
LTH 1237 \\                             
\vspace{0.3cm}                 
\today \\                                                    
\end{flushright} 
\vspace*{0.5cm}
\begin{center}                                                    
{\Large \bf Very low $x$ gluon density determined}\\
\vspace{0.5cm}
{\Large \bf   by LHCb exclusive $J/\psi$ data}\\
\vspace*{1cm}
 
C.A. Flett$^a$, A.D. Martin$^b$, M.G. Ryskin$^{b,c}$ and T. Teubner$^a$\\                                                   
%%%C.A. Flett$^a$, A.D. Martin$^c$, M.G. Ryskin$^{c,d}$ and T. Teubner$^a$\\                                                    
%%%S.P. Jones$^b$,                                                   
\vspace*{0.5cm}                                                    
$^a$  Department of Mathematical Sciences, University of Liverpool, Liverpool, L69 3BX, U.K.\\
%$^b$  Theoretical Physics Department, CERN, Geneva, Switzerland \\
$^b$ Institute for Particle Physics Phenomenology, Durham University, Durham, DH1 3LE, U.K. \\                                                   
$^c$ Petersburg Nuclear Physics Institute, NRC Kurchatov Institute, Gatchina, St.~Petersburg, 188300, Russia

\vspace*{1cm}                                                    
                                                    
\begin{abstract} 
The low $x$ behaviour of the gluon density $xg(x,\mu^2)$ at scale $\mu^2=2.4$ GeV$^2$ is determined using exclusive $J/\psi$ production data from HERA and LHCb within the framework of collinear factorisation at next-to-leading order (NLO).
%studied based on the description of exclusive $J/\psi$ production LHCb and HERA data within the complete collinear factorization NLO framework. 
It is shown that in the interval  $3\times 10^{-6} < x <10^{-3}$ the gluon distribution function grows as
$xg(x,\mu^2)\propto x^{-\lambda}$ with $\lambda=0.135\pm 0.006 $.  The impact this experimental data will have for the global parton distribution function (PDF) analyses in this low $x$ domain is quantified.
%The gluon density is found to be $xg=2.28\pm 0.06$ at $x=10^{-3}$, $\mu^2=2.4$ GeV$^2$. 
No indication in favour of parton density saturation is observed.
\end{abstract}

\vspace*{0.5cm}                                                    
                                                    
\end{center}

 \section{Introduction}
 \label{sec:intro}
 At  moderate values of $x$ the parton distribution functions (PDFs) of the proton are determined from global data with good accuracy by PDF analyses, see, for example, \cite{NNPDF,MMHT,CT14}. However in the low $x$ domain, with $x<10^{-3}$, there are practically no data to constrain the input parton densities. In this domain the gluon PDF dominates. The global PDF predictions are based simply on extrapolation using some more-or-less arbitrary ansatz for the input distributions. For this reason the uncertainties of the PDFs in the very low $x$ domain are huge.
 
There are two types of data which probe this domain.  The first is open charm production and the second is exclusive $J/\psi$ production -- both processes have been measured by the LHCb collaboration in the forward region. These processes probe mainly the gluon PDF at a rather low scale ${\cal O}(m_c)$ close to the input $Q_0$ values of the global PDF analyses. Here $m_c$ is the mass of the charm quark. Open charm production is experimentally more complicated to measure as it is extracted from $D$-meson production data, but the theoretical formalism is direct. On the other hand exclusive $J/\psi$ production is experimentally much cleaner, but the theoretical formalism needs care.  
 
 The charm and $D$-meson data~\cite{cc1} were used to restrict the uncertainty of the NNPDF gluon PDF in the low $x$ region in \cite{r7,r4,r5,r6,Bertone}.  However there are some inconsistencies in the energy and rapidity behaviour of the experimental results, which were discussed in \cite{Gauld,OMR}. The exclusive $J/\psi$ data are more consistent and have better accuracy than the inclusive $D$-meson cross section.
 
 Nevertheless, until now the $J/\psi$ data have not been used in global analyses due to theoretical complications. First, the $J/\psi$ cross section is driven by Generalised Parton Distributions (GPDs)
 and not directly by the conventional collinear PDFs. Second, the first calculations of the corresponding NLO coefficient functions revealed a huge scale uncertainty of the predictions \cite{Ivan,Diehl,J1610}. Both of these problems have been overcome.  
 
 The amplitude, $A$, for exclusive $J/\psi$ photoproduction may be written, using collinear factorization, in the form~\cite{Ivan}
\be
A = \frac{4 \pi \sqrt{4 \pi \alpha} e_q ( \epsilon_V^* \cdot \epsilon_\gamma)}{N_c} \left(\frac{ \langle O_1 \rangle_V}{m_c^3} \right)^{1/2} \int_{-1}^1 \frac{\mathrm{d}X}{X}\, \left[ C_g\left( X,\xi \right) F_g(X,\xi) + C_q(X,\xi) F_q(X,\xi) \right],
\ee
where we have suppressed the dependence on the renormalization and factorization scales, $\mu_R, \mu_F$, and on the invariant transferred momentum squared, $t$. Here, the non-relativistic QCD (NRQCD) matrix element $\langle O_1 \rangle_V$ describes the formation of the $J/\psi$ meson with $m_c$ the charm quark mass. The quantum numbers of the photon and the $J/\psi$ meson select the charge conjugation even exchange in the $t$-channel, so that only the charge conjugation even quark singlet and gluon GPDs, denoted $F_q$ and $F_g$ respectively, contribute. The quark and gluon coefficient functions $C_q$ and $C_g$ are known at NLO~\cite{Ivan}. 

\begin{figure} [t]
\begin{center}
\includegraphics[width=0.4\textwidth]{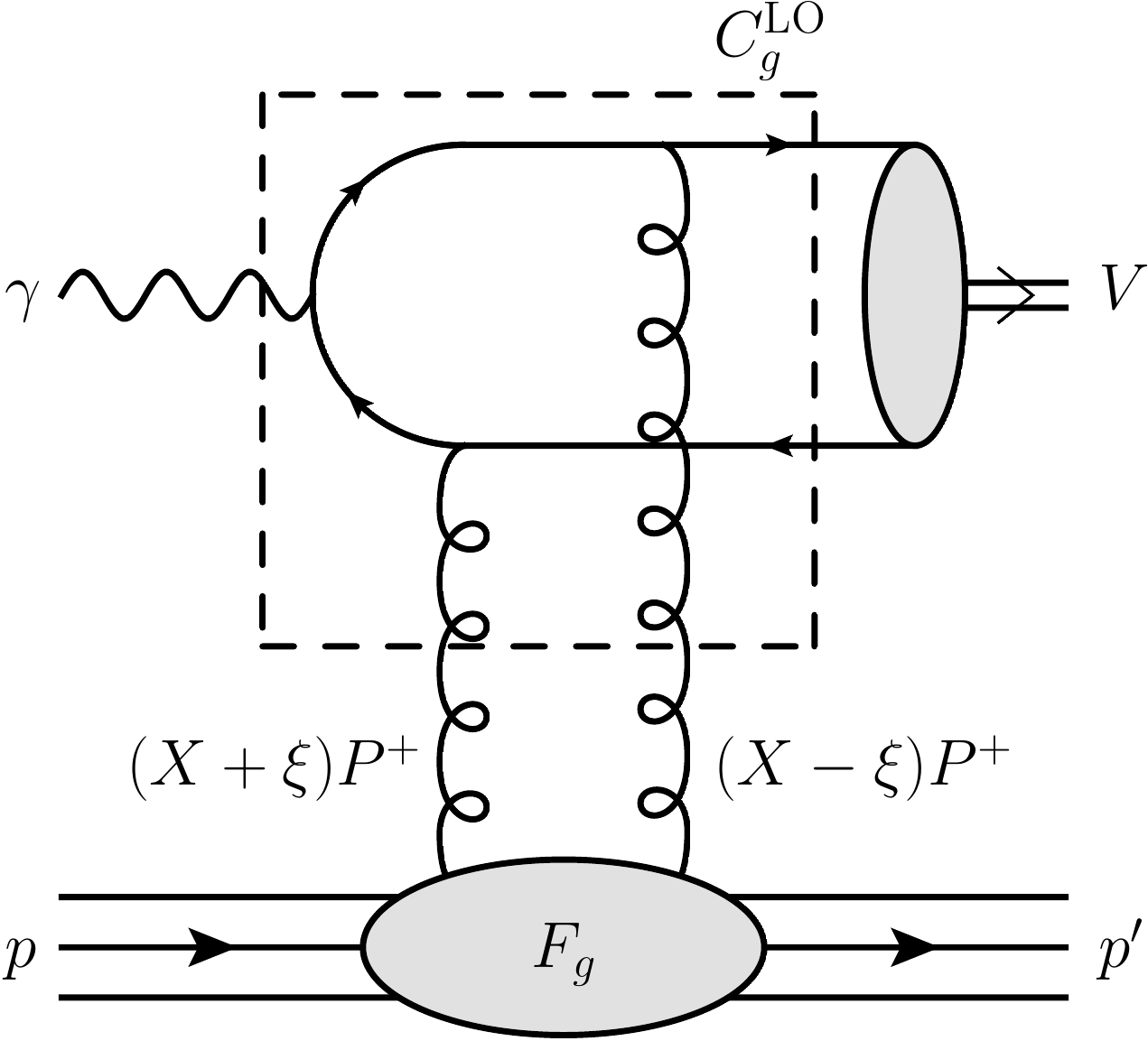}
\qquad
\includegraphics[width=0.4\textwidth]{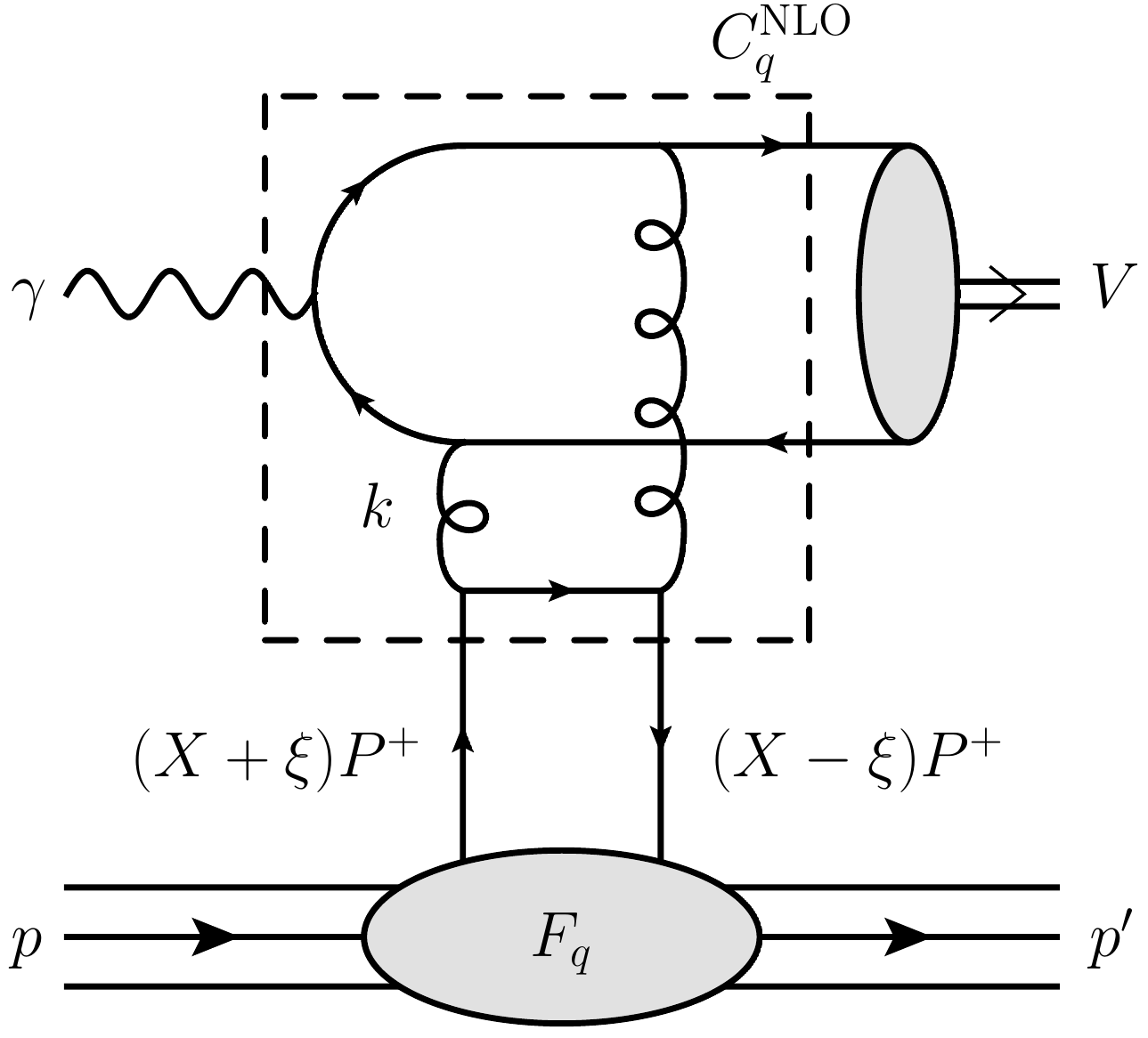}
\caption{\sf{  LO (left panel) and NLO (right panel) contributions to  $\gamma p \to V +p$, where $V=J/\psi$ (or $\Upsilon$).   Here the momentum  
  $P\equiv (p+p^\prime)/2$, with $k$ the loop momentum. 
  %and $l$ is the loop momentum. 
  Note that the momentum fractions of the left and right input partons are $x=X+\xi$ and $x'=X-\xi$ respectively; for the gluons coupled directly to the on-shell heavy quark pair,
  %for the upper gluons
  we have $x' \ll x$ and so $x\simeq 2\xi$.}}
\label{fig:f2}
\end{center}
\end{figure}

The kinematics of the process are displayed in Fig.~\ref{fig:f2}.
 %The amplitude, $A$, for exclusive $J/\psi$ photoproduction can be expressed, within collinear factorisation, as 
%\be
%A(\xi) \sim  \int_{-1}^1 \frac{\mathrm{d}X}{X}\, \left[ C_g\left( X,\xi \right) F_g(X,\xi) + C_q(X,\xi) F_q(X,\xi) \right],
%\ee
%where the $F_i$ and $C_i$ are, respectively, the GPDs and coefficient functions (known at NLO, see~\cite{Ivan}) and their dependence on the renormalisation and factorisation scales, $\mu_R^2, \mu_F^2$, has been suppressed, as well as that due to the invariant momentum transfer, $t$. 
%%%%%The projection of the outgoing quarks onto the final state $J/\psi$ meson is made within LO NRQCD with transition matrix element $\langle O_1 \rangle_V$. 
%%%%%%It is driven by the quasi-elastic semi hard scattering subprocess $\gamma p \rightarrow J/\psi~p$ for which the hard scattering kernels, $C_i$, are known at NLO. Due to the off forward kinematics and presence of skew parameter $\xi$, see Fig. 1, the process is susceptible to generalised PDFs (GPDs), denoted by $F_i$.  
%The kinematics of our set-up are shown in Fig.~\ref{fig:f2}. %%%%The forward PDFs are obtained from the LHAPDF6 interface [REF] and cast into the Shuvaev transform to generate GPDs, denoted by $F_i$. They are then convolved with the $Q_0$-subtracted coefficient functions, $C_i$, at NLO. [ NEED SAY SOMETHING ABOUT OUR USE OF SHUVAEV TRANSFORM IN RELATION TO OUR ANSATZ FOR LOW X GLUONS (WHICH ARE CURRENTLY NOT OFCOURSE IN LHAPDF FORMAT] ]
The partons carry momentum fractions $(X+\xi)$ and $(X-\xi)$ of the plus-component of the mean of the incoming and outgoing proton momenta, $P=(p+p^\prime)/2$, where the skewedness parameter $\xi$ is defined by
\be
\xi~=~ \frac{p^{+} - p^{\prime +}}{p^{+} + p^{\prime +}}~=~\frac{M^2_\psi}{2W^2-M^2_\psi},
%~<~10^{-3},
\ee
with $M_{\psi}$ the mass of the $J/\psi $ and $W^2$ the photon-proton energy squared for the $\gamma p\to J/\psi~p$ subprocess. It was shown that in the relevant region, $\xi~<~10^{-3}$, 
%\be
%\xi~=~ \frac{p^{+} - p^{\prime +}}{p^{+} + p^{\prime +}}~=~\frac{M^2_\psi}{2W^2-M^2_\psi}~<~10^{-3},
%\ee 
%depicted in Fig.~\ref{fig:f2}, 
the GPD functions can be related to the normal PDFs via the Shuvaev transform \cite {Shuv}. This relation is based on the fact that, due to the polynomial condition, the Gegenbauer moments of the GPDs are equal to the known Mellin moments of the non-skewed PDFs up to ${\cal O}(\xi)$ accuracy at NLO~\cite{Shuv,Nockles}. The GPD grids are generated from PDFs supplied on grids via the \texttt{LHAPDF} interface~\cite{LHAPDF}. 
%Here $M_{\psi}$ is the mass of the $J/\psi $ and $W^2$ is the photon-proton energy squared for the $\gamma p\to J/\psi~p$ subprocess.
  
 %The photon-proton centre of mass energy squared is given by $W^2=(q+p)^2$, where $q$ is the photon momentum. The asymmetry between the momentum fractions carried by the partons is parametrised by the skewness parameter,
%\be
%\xi = \frac{p^{+} - p^{\prime +}}{p^{+} + p^{\prime +}} = \frac{M_\psi^2}{2 W^2 - M_\psi^2}.
%\ee

%It was shown that in the relevant region of very small skewedness parameter $\xi$, 
%\be
%\xi~=~\frac{M^2_\psi}{2W^2-M^2_\psi}~<~10^{-3},
%\ee 
%depicted in Fig.~\ref{fig:f2}. The GPD functions can be related to the normal PDFs via the Shuvaev transform \cite {Shuv}. This relation is based on the fact that, due to the polynomial condition, the Gegenbauer moments of the GPD are equal to the known Mellin moments of the non-skewed PDF up to ${\cal O}(\xi)$ accuracy at NLO~\cite{Shuv,Nockles}. The GPD grids are generated from PDFs supplied on grids via the \texttt{LHAPDF} interface~\cite{LHAPDF}. Here $M_{\psi}$ is the mass of the $J/\psi $ and $W^2$ is the photon-proton energy squared for the $\gamma p\to J/\psi~p$ subprocess. 
%%%%The skewedness parameter, $\xi$, is shown in Fig.~\ref{fig:f2}. 
%The relation is based on the fact that, due to the polynomial condition, the Gegenbauer moments of the GPD are equal to the known Mellin moments of the non-skewed PDF up to ${\cal O}(\xi)$ accuracy \cite{Shuv,Nockles}. 

The second problem, that concerns the strong dependence on the factorization scale observed in the low $x$ region, was essentially removed by subtracting the low $k_t<Q_0$ contribution from the NLO coefficient functions. This subtraction is needed to avoid the double counting between the NLO coefficient function and the contribution hidden in the input PDF \cite{J1610}.  In addition the double log terms, [$\alpha_s\ln \mu^2_F \ln (1/x)]^m$, can be resummed in the leading order term by choosing the optimum scale $\mu_F=m_c=M_{\psi}/2$ for our process~\cite{Jdl}. The NLO amplitude $A(\mu_f)$, with factorisation scale $\mu_f$, can be written schematically in the form 
\be
\label{2}
A(\mu_f)~=~C^{\rm LO} \otimes {\rm GPD}(\mu_F)~+~C^{\rm NLO}_{\rm 
rem}(\mu_F)\otimes{\rm GPD}(\mu_f).
\ee
With the choice $\mu_F=M_{\psi}/2$, the remaining NLO coefficient function, $C^{\rm NLO}_{\rm 
rem}(\mu_F)$, does not contain terms enhanced by ln$(1/x)\simeq\ \ln(1/\xi)$. 
%and has a weaker dependence on the (residual) factorisation scale, $\mu_f$.

The approach was described in more detail in \cite{PRD} where it was shown that the HERA data on diffractive $J/\psi$ photoproduction \cite{HERA} with energies corresponding to $x>10^{-3}$ are well described using the present global gluons.\footnote{We should also mention the possibility of relativistic
corrections to the NRQCD matrix element that we use in our approach. Recall that,
strictly speaking, if we were to include relativistic corrections, see for example~\cite{Finland}, then we must
simultaneously account for the higher, $ c \bar{c} + g$, Fock component of the
$J/\psi$ wave function. As was shown
in~\cite{Hood}, these two corrections
largely cancel  each other, leading to a final correction of the order of a
few percent provided that the NRQCD matrix element is normalized to the leptonic decay width,
$J/\psi\to l^+l^-$, and the charm quark mass is chosen to be
$m_c=M_{\psi}/2$, as is kept in the present paper. Note also that the
correction to the NRQCD matrix element changes the normalization of the $J/\psi$ cross section but does not
affect the $x$ (or $W$) behaviour of the low-$x$ gluon. The fact that at $x \gapproxeq 0.001$ the data are well described by the existing global gluons is an argument in favour of the correct
normalization, that is, in favour of small relativistic corrections to
our approach.}  This demonstrated the efficiency of the method, which will be used in the present note to extract the behaviour of the gluon in the low $x$ region ($x<10^{-3}$) from the exclusive $J/\psi$ LHCb data \cite{LHCb} (as well as HERA photoproduction data that lie in this region).

As was shown in \cite{PRD}, after the $k_t<Q_0$ subtraction the quark contribution to this process is negligibly small in this $x$ region. Thus we determine just the gluon PDF and use the quark PDF from the existing global fits.

Of course, at the moment, global PDF analyses are performed to NNLO accuracy. However, as a first step, we start fitting the $J/\psi$ data at NLO.  In the future this approach can be extended to NNLO.\footnote{This would require knowledge of the 2-loop hard scattering coefficient function.}
%The $Q_0$ subtraction part is simpler and based on the 2-loop form therein.}.

The outline of the paper is as follows.  In Section 2 we describe the
ansatz that we will use to parametrize the NLO gluon PDF in the collinear factorization scheme in the low $x$ domain, $x<0.001$. In Section 3, after a brief discussion of the exclusive $J/\psi$ data, we describe how we determine the low $x$ gluon directly from the data. 
%The results are presented in Section 3. 
 In Section 4, we compare the results we find for the low $x$ gluon with those obtained by reweighting the NNPDF gluon using the $D$-meson  LHCb data.
Finally, in Section 5, we provide a reweighting of the NNPDF3.0 gluon via the exclusive $J/\psi$ data and compare and contrast this with the gluon obtained from the above alternative approaches. 
%is devoted to a short discussion of alternative analyses. 
Our conclusions are briefly summarized in Section 6.

\section{Ansatz for the low $x$ gluon} 
It was demonstrated in \cite{PRD} that the diffractive $J/\psi$ cross section is driven by the Generalised Parton Distributions,   GPD$(X+\xi,X-\xi)$, of the gluon with $X\simeq \xi$, see Fig.~\ref{fig:f2}. That is, to describe the LHCb data, we effectively need the gluon in the region of low $x\simeq X+\xi $ only. So it is sufficient to parametrize the gluon in the region $x<10^{-3}$. On the other hand the Shuvaev transform, that relates the GPD to the conventional collinear gluon PDF, includes an integral over the whole $x<1$ interval. Moreover, the transform was derived assuming that the gluon had a smooth analytical behaviour  with the property that $g(x)\to 0$ as $x\to 1$.  In order to satisfy these requirements we choose the following ansatz for the conventional gluon PDF,
\be
xg(x,\mu_0^2)~=~C ~xg^{\rm global}(x,\mu_0^2)~+~(1-C)~xg^{\rm new}(x,\mu_0^2)
\label{eq:az}
\ee
\be
\label{c}
{\rm with}~~~~~C~=~\frac{x^2}{x^2+x^2_0}~,
\ee
and where $xg^{\text{global}}$ is the value of the gluon PDF obtained in a global PDF analysis. The simplest low $x$ form for the gluon would be
\be
xg^{\rm new}(x,\mu_0^2)~=~nN_0~(1-x)~x^{-\lambda},
\label{eq:pow}
\ee 
where the normalization factor $N_0$ is chosen so that for $n=1$ the gluon PDF has 
%a smooth analytical matching at $x=x_0$,
the matching at $x=x_0$,
\be
x_0g^{\rm new}(x_0,\mu_0^2)~=~x_0g^{\rm global}(x_0,\mu_0^2).
\label{eq:match}
\ee
%where $xg^{\rm global}$ is the value of the gluon PDF obtained in a global PDF analysis. 
The factor $n$ in (\ref{eq:pow}) is close to 1. It allows the possibility of
matching to a global gluon whose normalization differs from $N_0$ but still lies
 within the global gluon error band at $x=x_0$.  
The factor $(1-x)$ in (\ref{eq:pow}) provides the vanishing $xg\to 0$ as $x\to 1$.\footnote{Note that this factor was added to satisfy the formal conditions for the validity of the Shuvaev transform. Practically, the results do not depend on the behaviour of the gluon at relatively large $x$. The corresponding effects are not visible in our Figs.~\ref{fig:2},~\ref{fig:xg},~\ref{fig:6}.} Due to the smooth form of $C$ in (\ref{c}) the complete distribution
(\ref{eq:az}) does not violate analyticity even for $n\neq 1$.

Alternatively, in order to compare our present collinear determination of $xg^{\rm new}$ with
an earlier determination of the low $x$ gluon obtained in the $k_t$ factorization
approach \cite{Jones}, we also use an ansatz inspired by the double logarithm approximation, 
\be
\label{eq:kt}
xg^{\rm new}(x,\mu_0^2)~=~nN_0~
(1-x)~x^{-a}\left(\frac{\mu_0^2}{q^2_0}\right)^{-0.2}~{\rm
exp}\left[\sqrt{16(N_c/\beta_0)\ln (1/x)\ln G)}\right]
\ee
\be
{\rm with}~~~~G~=~\frac{\ln (\mu_0^2/\Lambda^2_{\rm QCD})}{\ln (q_0^2/\Lambda^2_{\rm
QCD})}~,
\ee
where the parameter $a$ now plays the role of $\lambda$. Here, with three light quarks $(N_f=3)$ and $N_c=3$ we have $\beta_0=9$. We take $\Lambda_{\text{QCD}}=200\,\text{MeV}$ and $q_0^2 = 1\,\text{GeV}^2$, as in \cite{Jones}, with $\mu_0^2 = 2.4\,\text{GeV}^2$ fixed.  The exponent in
(\ref{eq:kt}) resums, to all orders in $m$, the double logarithmic terms $(\alpha_s\ln(1/x)\ln \mu^2)^m$
and hence we find that, to good accuracy, we reproduce the NLO DGLAP low $x$ evolution
in the interval of $Q^2$ from 2 to about 30 GeV$^2$. Therefore this parametrization
can be used to describe $\Upsilon$ photoproduction data as well.

%\be
%\label{eq:pow}
%xg^{\rm new}(x,\mu_0^2)~=~N~(1-x)x^{-\lambda}
%\ee 
%where $N=N_0n$ and the normalization factor $N_0$ is chosen so that at $n=1$ the gluons have a smooth analytical matching at $x=x_0$ 
%\be
%x_0g^{\rm new}(x_0,\mu_0^2)~=~x_0g^{\rm global}%(x_0,\mu_0^2).
%\label{eq:match}
%\ee 
%Second factor $n$ is close to 1. It provides the possibility for some variations of the low-$x$ gluons normalization within the global gluons error band  
%while the factor $(1-x)$ provides the vanishing $xg\to 0$ as $x\to 1$. Note that thank to the smooth form of $C$ (\ref{c}) the complete distribution (\ref{eq:az}) does not violate the analyticity even for $n\neq 1$.

%Next, in order to compare our present collinear determination of $xg^{\rm new}$ with an earlier determination of the low $x$ gluon obtained in the $k_t$ factorization approach \cite{Jones}, we also use the ansatz 
%\be
%\label{eq:kt}
%xg^{\rm new}(x,\mu_0^2)~=~N(1-x)x^{-a}%\left(\frac{\mu_0^2}{Q^2_0}\right)^{-0.2}~{\rm exp}%\left[\sqrt{16(N_c/\beta_0)\ln (1/x)\ln G)}\right]
%\ee
%\be
%{\rm with}~~~~G~=~\frac{\ln (\mu_0^2/\Lambda^2_{\rm QCD})}{\ln (Q_0^2/\Lambda^2_{\rm QCD})}
%\ee
%where the single parameter $a$ now plays the role of $\lambda$.  The exponent in (\ref{eq:kt}) resums the double logarithmic terms $(\alpha_s\ln(1/x)\ln \mu^2)^n$ and hence to good accuracy we find that we reproduce the NLO DGLAP low $x$ evolution in the interval of $Q^2$ from 2 to about 30 GeV$^2$. Therefore this parametrization can be used to describe $\Upsilon$ photoproduction data as well.

\section{Determination of the low $x$ gluon from $J/\psi$ data}

%Figs.~\ref{fig:2} and \ref{fig:3} 
Here, we show the results of our fits to $J/\psi$ photoproduction data for $x<10^{-3}$, using an ansatz for the gluon PDF as described in eqns.~(\ref{eq:az})--(\ref{eq:match}). The matching is made at $x_0 = 10^{-3}$ using the gluon PDF from three NLO parton global analyses,~NNPDF3.0~\cite{NNPDF},  MMHT14~\cite{MMHT} and CT14~\cite{CT14}.
%which matches the gluon PDF to that of three global NLO parton sets \cite{NNPDF,MMHT,CT14} at $x_0=10^{-3}$ as described in eqs. (\ref{eq:az})--(\ref{eq:match}), in order to determine the gluon PDF from data for $x<10^{-3}$. 
Due to the small contribution of the quark sector at NLO to the $J/\psi$ cross section~\cite{PRD}, we do not attempt to fit the quark PDFs but only the gluon PDF around its input scale. The quark PDFs obtained in the global NLO analyses are therefore used for all $x$.

\subsection{The exclusive $J/\psi$ data from LHCb}
The LHCb experiment, by design, does not directly measure the cross section for $J/\psi$ {\it photoproduction} but instead that for exclusive $pp \to p + J/\psi + p$ \cite{LHCb}. The experiment is unable to tag forward protons accompanying the $J/\psi$ so instead only the rapidity of the $J/\psi$ is measured. Events are selected by ensuring a large rapidity gap on both sides of the $J/\psi$ measurements, where the transverse momentum of the $J/\psi$ is small, and assumed to correspond to {\it exclusive} reactions.  The lack of forward proton tagging means it is also not possible to determine which of the two protons emitted the photon.
%%%%\begin{figure}[h]
%%%%\begin{center}
%\vspace*{-1.cm}
%%%%\includegraphics[scale=0.55]{Fig1EPSFINAL.pdf}
%\vspace*{-4cm}
%%%%\caption{\sf{The two diagrams describing exclusive $J/\psi$ production at the
 %%%% LHC. The vertical lines represent two-gluon exchange. Diagram (a), the $W_+$ component, is the major contribution to the $pp \to p+J/\psi +p$ cross section for a $J/\psi$ produced at large rapidity $Y$. Thus such data allow a probe of very low $x$ values, $x\sim M_{\psi} {\rm  exp}(-Y)/\sqrt{s}\,$; recall that for two-gluon exchange we have $x\gg x'$.   The $q_T$ of the photon is very small and so the photon can be considered as a real on-mass-shell particle.}}    
\begin{figure}[h]
\begin{center}
%\vspace*{-1.cm}
\includegraphics[scale=0.8]{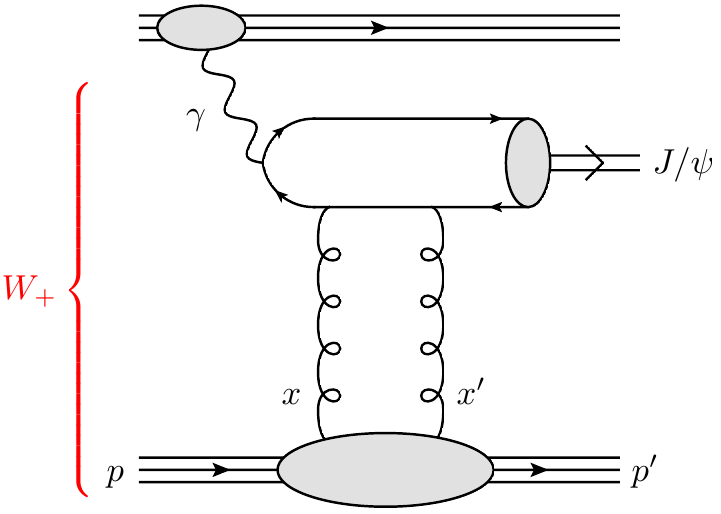}
\qquad
\includegraphics[scale=0.8]{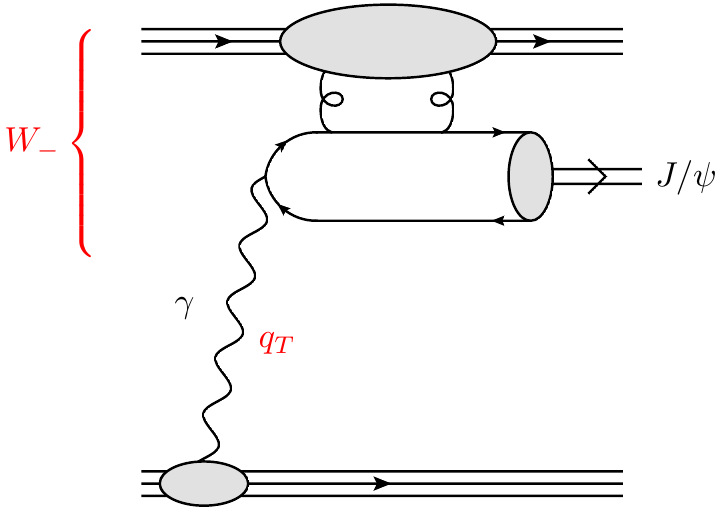}
%\includegraphics[scale=0.55]{Fig1EPSFINAL.pdf}
%\vspace*{-4cm}
\caption{\sf{Two leading-order (LO) diagrams describing exclusive $J/\psi$ production at the
  LHC. 
  %The vertical lines represent two-gluon exchange. 
  The left diagram,
  the $W_+$ component, is the major contribution to the $pp \to p+J/\psi
  +p$ cross section for a $J/\psi$ produced at large rapidity $Y$. Thus
  such data allow a probe of very low $x$ values, $x\sim M_{\psi} {\rm
    exp}(-Y)/\sqrt{s}\,$; recall that for two-gluon exchange we have
  $x\gg x'$.   The $q_T$ of the photon is very small and so the photon can be considered as a real on-mass-shell particle.}}    
\label{fig:LHCb1}
\end{center}
\end{figure}
The ultraperipheral amplitude for a given $J/\psi$ rapidity is then generally the sum of two photoproduction amplitudes with different $W^2$, depending on which proton emitted the photon and which was the target, see Fig.~\ref{fig:LHCb1}. The interference contribution is suppressed as the photon’s transverse momentum, $q_T$, is much smaller than that of the proton exchanging the gluons. 
%(shown by the double vertical lines in Fig.~\ref{fig:LHCb1}).  
The contribution corresponding to the right graph, with a smaller photon-proton energy $W_-$ , comes from relatively large $x$, and can be subtracted using  the existing description of HERA data. The cross section for $J/\psi$ photoproduction at the large energy, $W_+$, may therefore be extracted from the LHCb measurements. 

Additionally, at the LHC, there is a non-negligible probability of additional soft interactions between the two colliding protons that can result in secondary particles polluting the rapidity gaps used to select the exclusive events. This will suppress the number of events deemed exclusive and therefore one must account for the gap survival probability, $S^2 < 1$, to have no such additional interaction. The value of $S^2$ depends on the $pp$ collider energy and the partonic energy $W$. The values
of $S^2(W)$ as a function of $W$ were calculated using the eikonal model~\cite{KMR74} which well describes the data for the differential $\mathrm{d}\sigma(pp)/\mathrm{d}t$ cross section and low-mass diffractive dissociation. The details of the procedure to extract $\sigma(\gamma p\to J/\psi+p)$ at large $W_+$ energies is described in~\cite{Jones}.
%In Fig.~\ref{fig:2} 
We use the low $x$ LHCb ``data" points obtained in this way by the LHCb collaboration~\cite{LHCb}. 

%Another point is that in dealing with proton-proton interactions we must account for the possibility of an additional soft interaction between the two colliding protons. This interaction will generate new secondaries which  will populate the rapidity gap and  destroy the exclusivity of the event. The probability to have no such additional interaction is called the gap survival probability $S^2<1$.
%The value of $S^2$ depends on the $pp$ collider energy and the partonic energy $W$. The values
%of $S^2(W)$ as a function of $W$ were calculated using the eikonal model~\cite{KMR74} which well describes the data for the differential $\mathrm{d}\sigma(pp)/\mathrm{d}t$ cross section and low-mass diffractive dissociation. The details of the procedure to extract $\sigma(\gamma p\to J/\psi+p)$ at large $W_+$ energies is described in reference~\cite{Jones}.
%In Fig.~\ref{fig:2} we plot the low $x$ LHCb ``data" points obtained in this way by the LHCb collaboration~\cite{LHCb}. 
%(i.e. taken from) in~\cite{Jones}.   ~\cite{LHCb}.

\begin{figure} [t]
%%\vspace{-1.3cm}
\begin{center}
%\text{Uncertainty of global gluon at $x=x_0=0.001$\,\,\,\,\,Uncertainty of gluon at $x=0.001$ from fit to HERA+LHCb $J/\psi$ data} \par\medskip
\includegraphics[scale=0.7]{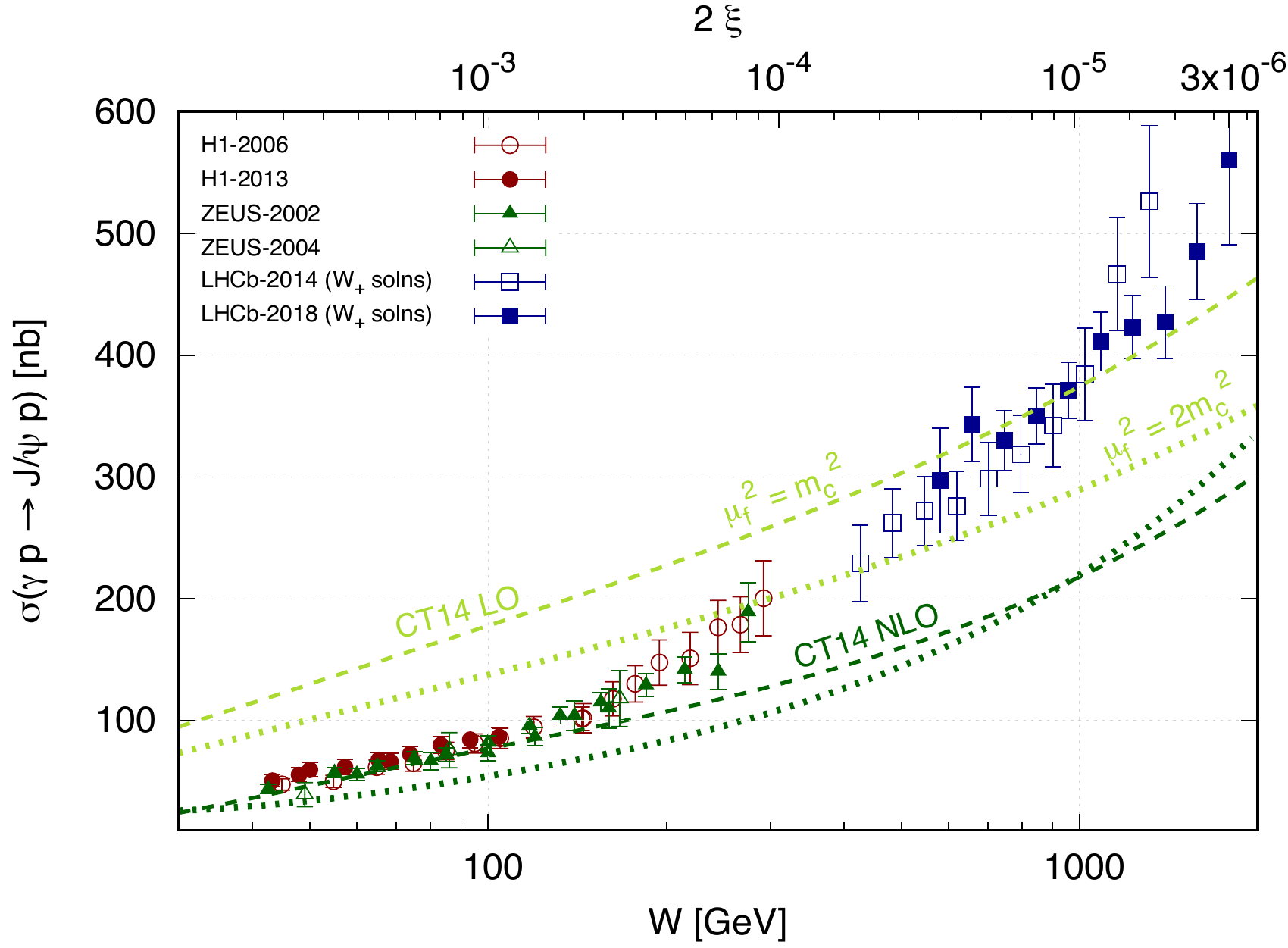}
%%\includegraphics[width=1.2\textwidth]{xg001.pdf}
%%\vspace{-6.0cm}
\caption{\sf{ LO and NLO cross section predictions obtained using the central values of the existing global partons from~\cite{CT14}. Dashed (dotted) lines correspond to the scale choices $\mu_f^2=\mu_R^2=m_c^2$ ($\mu_f^2 = \mu_R^2 = 2m_c^2$) with $\mu_F = Q_0 = m_c$ fixed. }}
\label{new}
\end{center}
\end{figure}

\begin{figure} [t]
%%\vspace{-2.0cm}
\begin{center}
\includegraphics[scale=0.7]{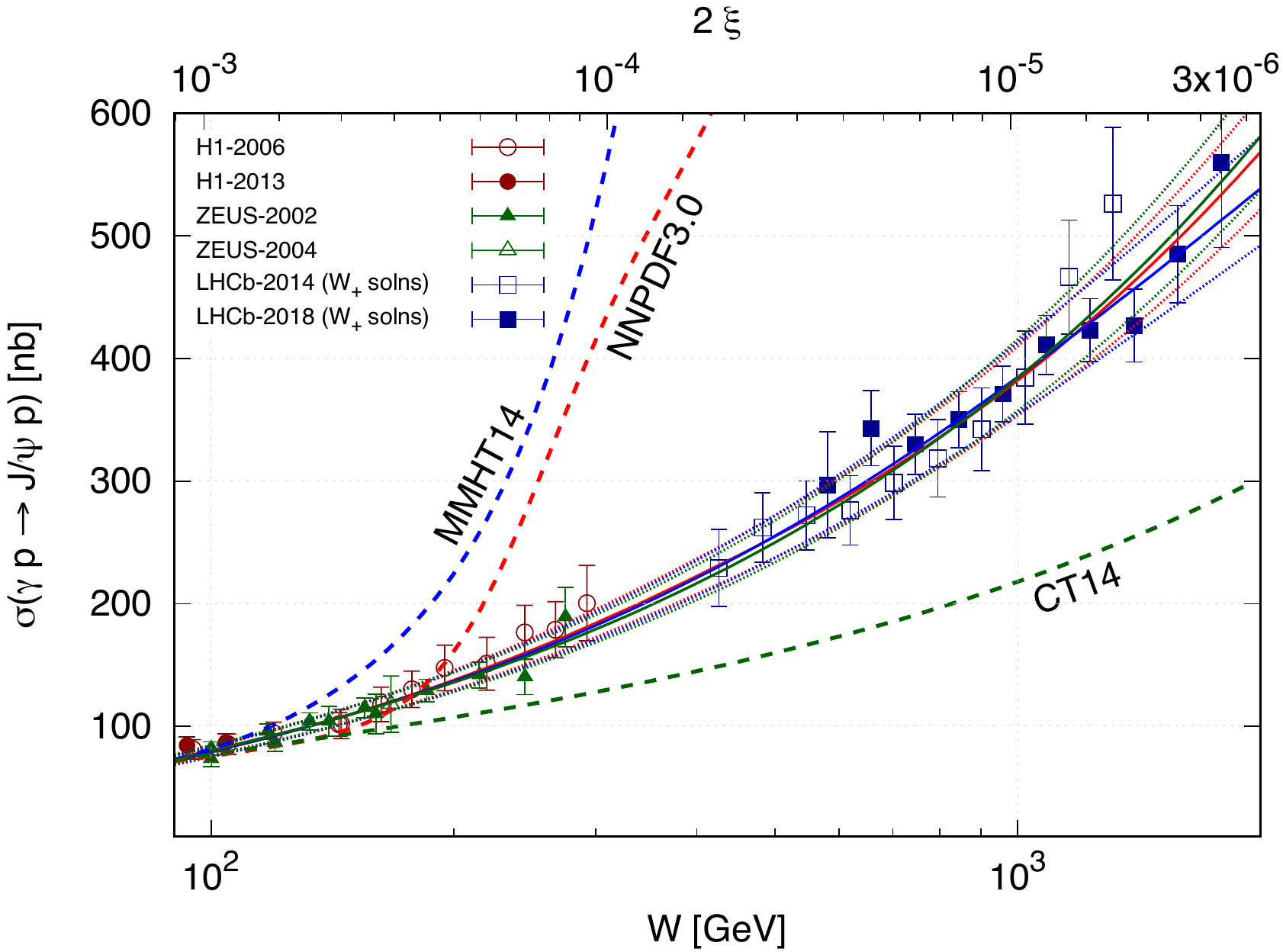}
%%\includegraphics[width=1.3\textwidth]{Fig3.pdf}
%%\vspace{-5.0cm}
\caption{\sf{The description of the $J/\psi$ photoproduction HERA~\cite{HERA} and LHCb~\cite{LHCb} data based on using the central value of the global gluon PDF from the three global parton analyses~\cite{NNPDF,MMHT,CT14} for $x>0.001$. 
%given by the three global parton analyses \cite{NNPDF,MMHT,CT14}. 
The red, blue and green solid and dotted lines show the spread of descriptions based on the power fit, using the $\pm\,1 \sigma$ errors for the parameters. We also show by dashed lines the cross section predictions obtained using the 
%what would happen if 
current central values of the global gluons for all $x$.}}
%were to be  extrapolated into the low $x$ domain (green, blue and black dashed lines).  
%The solid grey and black lines are, respectively, the cross section predictions obtained using the central value of the NNPDF3.1 (with low $x$ BFKL resummation) global partons constrained by the $D$ meson LHCb data, REF, and NNPDF3.0 global partons supplemented with $D$ meson constraints, REF. (PROBABLY PUT THESE LINES IN ANOTHER SEPERATE CROSS SECTION PLOT}}
\label{fig:2}
\end{center}
\end{figure}

%NLO gluon fit from the J\psi data
\subsection{Description of the $J/\psi$ data}

The set-up of the cross section prediction follows~\cite{PRD}. Only the imaginary part of the amplitude is computed using eq.~(1). In this way we need only the GPDs in the DGLAP region $|X| > \xi$.
%where we expect an absence of Mellin $N$-poles in the complex angular momentum plane and validity of the Shuvaev transform. 
In the ERBL region, $|X| < \xi$, the imaginary part of the coefficient function is zero.  The real part is then restored via a dispersion relation, which in the high energy limit (for the even signature amplitude) can be written in the simplified form \cite{Re}
\be
\frac{{\rm Re}A}{ {\rm Im}A}~~=~~{\rm tan}\left(\frac{\pi}{2}~\frac{\partial(\ln{\rm Im}A/W^2)}{\partial(\ln W^2)}\right).
\ee
Next, we use NRQCD to describe the formation of the $J/\psi$ wave function. We project the open heavy $c \bar{c}$ quark pair onto the colour singlet configuration with the corresponding transition matrix element $\langle O_1\rangle _V$, which is fixed by the experimentally measured leptonic decay width of the $J/\psi$. The exclusive final state requires a colourless high energy scattering (modelled by the two-gluon exchange) and does not allow for an octet contribution, as this would populate the rapidity gap and destroy the exclusivity of the final state.  

Note that actually we calculate the value of Im$A$ at $t=0$ and then restore the
total $\gamma p\to J/\psi+p$ cross section assuming an exponential $t$
behaviour with a slope
$$ B=4.9+4\alpha'_P\ln(W/W_0) ~~\mbox{GeV}^{-2}$$
with $W_0=90$ GeV and $\alpha'_P=0.06$ GeV$^{-2}$. This parametrisation grows more slowly with $W$ than the formula used by H1 \cite{HERA}, but is still compatible with the HERA data. We have chosen the slope parameter $\alpha'_P$ to be compatible with Model 4 of \cite{Diffraction} which fits a wider variety of data.

To set the scene, we first use eq.~(1) at LO and NLO to generate and compare cross section predictions using the existing LO and NLO partons from~\cite{NNPDF,MMHT,CT14}, respectively, for the $x$-range where we have used exclusive $J/\psi$ data from H1, ZEUS and LHCb.
%{\it all} $x$, respectively. 
In this way, we are able to quantify the scale dependence of the theoretical prediction as well as the size of the NLO result relative to the LO one. In Fig.~\ref{new}, we show such a comparison using CT14 partons~\cite{CT14}. Our choice of scales is explained in~\cite{Jdl}.  The NLO scale variation is smaller than that at LO and a better description of the HERA data is obtained with the NLO result. The plot emphasises that in the region where the current PDFs are well constrained, it is still crucial to use the NLO description. It is reassuring and non-trivial that our NLO prediction, with the `optimum' scale choice, agrees well with the HERA data.

We now determine the low-$x$ gluon by performing a two-parameter ($\lambda$ and $n$, as defined in eq.~(\ref{eq:pow})) fit of all the $\sigma(\gamma p\to J/\psi+p)$ LHCb and HERA data with $x<0.001$ using, as input, NLO parton PDFs from~\cite{NNPDF,MMHT,CT14}.
%determined in three different global parton analyses: 
The results are shown in Table \ref{tab:1} and Fig. \ref{fig:2}.

\begin{table}[ht]
\centering
\begin{tabular}[t]{lcccc}
\toprule
%%&$\lambda$&$n$&$\chi_{\text{min}}^2$\\
&$\lambda$&$n$&$\chi_{\text{min}}^2$&$\chi_{\text{min}}^2/\text{d.o.f}$\\
\midrule
%NNPDF3.0&0.153&0.95&15.4\\
%NNPDF3.0&0.153&0.95&44.98&1.05\\
NNPDF3.0&0.136&0.966&44.51&1.04 \\
%%MMHT14&0.158&1.05&15.3\\
%%MMHT14&0.158&1.05&47.38&1.10\\
MMHT14&0.136&1.082&47.00&1.09\\
%%CT14&0.150&0.92&15.5\\
%%CT14&0.150&0.92&48.29&1.12\\
CT14&0.132&0.946&48.25&1.12\\
\bottomrule
\end{tabular}
\caption{\sf{The values of $\lambda$ and $n$ obtained from fits to the $J/\psi$ data using three sets of global partons. The respective values of the total $\chi_{\text{min}}^2$ (and $\chi_{\text{min}}^2/\text{d.o.f}$) for 45 data points are also shown.}} 
%giving e.g. a $\chi_{\text{min}}^2/\text{d.o.f} = 1.05$ for NNPDF3.0.}}
%\caption{\sf{The values of $\lambda$ and $n$ obtained from fits to the $J/\psi$ data using three sets of global partons.
%using the central values of the global gluons at $x=0.001$. 
%The respective values of the total $\chi_{\text{min}}^2$ for 45 data points are also shown giving e.g. a $\chi_{\text{min}}^2/\text{d.o.f} = 0.36$ for NNPDF3.0.}}
\label{tab:1}
\end{table}%

%%\begin{table}[h!]
%%\begin{center}
%%\renewcommand\arraystretch{1.15}
%%\begin{tabular}{|c|c|c|c|}
%%\hline
%%global set &$\lambda$ & $n$&$\chi^2$ \\
%%\hline
%%NNPDF3.0& 0.153 & 0.95& 15.4\\
%%MMHT14  & 0.158& 1.05&15.3\\
%%CT14 &    0.150& 0.92&15.5 \\
%%\hline
%%\end{tabular}
%%\caption{\sf{The values of $\lambda$ and $n$ obtained from fits to the $J/\psi$ data using three sets of global gluons.
%using the central values of the global gluons at $x=0.001$. 
%%The respective values of $\chi^2$ for 45 data points are also given.}}
%%\label{tab:1}
%%\end{center}
%%\end{table}
The respective values of the
$\chi_{\text{min}}^2$ statistic were calculated accounting for the bin-to-bin correlated errors within each individual experimental data set as well as uncorrelated errors. The covariance matrix was constructed, and iterated, according to the `$t_0$ prescription' as outlined in \cite{TO}. We use all HERA data points~\cite{HERA} with $W>100$ GeV and all LHCb~\cite{LHCb} data points.

For the  ZEUS  2002  and  2004  data  sets  \cite{HERA} we  allow  for  a  fully  correlated  $6.5\%$  normalisation
error.  For the H1 2006 data set \cite{HERA} we include a fully correlated $5\%$ normalisation error while for the H1 2013 data set \cite{HERA} we use
the full covariance matrix as provided by H1.  For the LHCb 2014 data \cite{LHCb} we allow for a fully
correlated $\sim7\%$ normalisation error. Finally, for the  LHCb 2018 data \cite{LHCb}, we use the covariance matrices supplied by the collaboration as well as a fully correlated normalisation error of $\sim4\%$.
%using the full diagonal covariance
%matrix of correlated errors accounting for 
% all the HERA data points~\cite{HERA} with $W>100$ GeV
%  and for the all LHCb~\cite{LHCb} data points [NEEDS REWORDED AND FURTHER INPUT USING FULL COVARIANCE MATRIX]. \\
  
  %{\bf \Large  Chris, Thomas, Stephen --}\\
 % {\bf Do we want (?) to write a bit more about the $chi^2$ calculations like this was done in 1611.03711 --" Additionally, we improve our fitting procedure to allow also for bin-to-bin
%correlated  errors  within  each  individual  data  set  as  well  as  uncorrelated  errors.   For  each  of
%the  ZEUS  2002  and  2004  data  sets  [5,  6]  we  allow  for  a  fully  correlated  6.5\%  normalisation
%error.  For the H1 2006 data set [7] we include a fully correlated 5\% normalisation error (also
%between the photoproduction and electroproduction data).  For the H1 2013 data [8] we use
%the full covariance matrix as provided by H1.  For the LHCb 2014 data [9] we allow for a fully
%correlated 7\% normalisation error.  The LHCb 2013 data [4] are superseded by the 2014 data
%and  are  not  included.   For  the  preliminary  LHCb  2016  data  [1]  we  take  the  fully  correlated
%normalization error of 7\%."}

\begin{figure} [t]
%%\vspace{-1.3cm}
\begin{center}
%\text{Uncertainty of global gluon at $x=x_0=0.001$\,\,\,\,\,Uncertainty of gluon at $x=0.001$ from fit to HERA+LHCb $J/\psi$ data} \par\medskip
\includegraphics[scale=0.7]{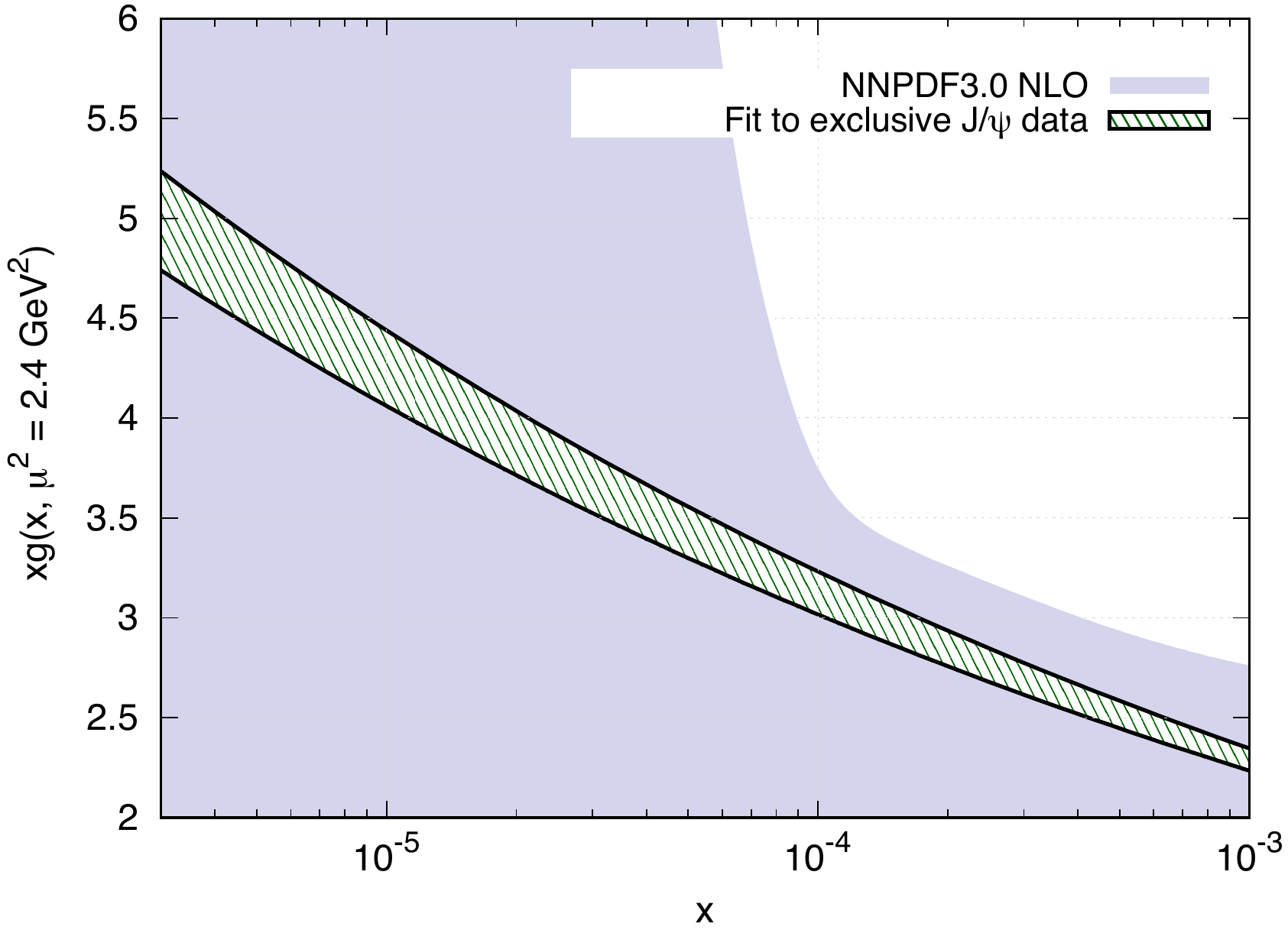}
%%\includegraphics[width=1.2\textwidth]{xg001.pdf}
%%\vspace{-6.0cm}
\caption{\sf{The cross-hatched region shows the range of behaviour of the low $x$ NLO gluon determined by fitting to exclusive $J/\psi$ data using ansatz (\ref{eq:pow}) with $xg^{\rm global}$ taken from NNPDF3.0~\cite{NNPDF}, MMHT14~\cite{MMHT} or CT14~\cite{CT14} parton sets.
%%The low $x$ gluon PDF $xg(x,\mu^2)$ with $\mu^2=2.4\, \text{GeV}^2$ determined from the fit to the exclusive $J/\psi$ data. 
%The red, blue and green solid lines show the central values obtained from matching to NNPDF3.0, MMHT14 and CT14 global partons at $x=0.001$. 
%%The shaded area is obtained in the manner as described within the text.  
}}
\label{fig:xg}
\end{center}
\end{figure}

%The three %different descriptions of the data are shown in Fig.~\ref{fig:2}. 
%%The extracted gluons based on the power fit description are shown in Fig. 4. 
The description of the exclusive $J/\psi$ cross section is shown in Fig.~\ref{fig:2}, while the gluons extracted  from the $J/\psi$ data at $\mu^2=2.4$ GeV$^2$ and $x<0.001$ are shown in Fig.~\ref{fig:xg}. The error bands are obtained by sampling over the two parameters within their individual 1$\sigma$ standard deviations, accounting for their correlation. The hatched green band in Fig.~\ref{fig:xg} in addition accounts for the 
%as well as the 
%The error band accounts for the full correlation matrix between $n$ and $\lambda$ parameters and the 
uncertainty due to the choice of the
global (NNPDF3.0, MMHT2014 or CT14) partons. 
As is seen from Fig.~\ref{fig:xg}, the resulting gluon at very small $x$ shows no hint of the onset of saturation. 
%[KEEP THIS LAST SENTENCE FOR THE DISCUSSION LATER]

 Starting from three different sets of global partons, we obtain practically the same low $x$ gluons
with the same quality ($\chi_{\text{min}}^2$) of the description.
The typical errors are $\pm 2.5$\% for the normalization and $\pm 4$\% for $\lambda$. We see from Fig.~\ref{fig:2} that the simple two-parameter form of the gluon density provides an excellent description of the $J/\psi$ data in the fitted $x<10^{-3}$ region, irrespective of which global parton set is used. In fact, the three descriptions only visibly differ for $x<10^{-5}.$ Note that the observed hierarchy of central cross section predictions at $x \sim 3 \times 10^{-6}$ differs from that expected given the power behaviours in Table 1. We have checked that this is due to the small $x$ and small scale quark behaviour of the global sets. 

% Recall that the description is based solely on the gluon, which is by far the dominant PDF in this domain. As expected, for larger $x$ we see that the simple description starts to become insufficient.  
Figure~\ref{fig:2} also shows the cross section predictions obtained using the central values of the gluon from the global parton sets extrapolated into the low $x$ region.  Clearly here the global analyses have no predictive power and in each case they have huge uncertainty bands (shown in Fig.~\ref{fig:xg} for NNPDF3.0 only) which cover the (unfitted) $J/\psi$ data. The value of including the $J/\psi$ data is apparent.

\begin{figure} [t]
%%\vspace{-1.3cm}
\begin{center}
%\text{Uncertainty of global gluon at $x=x_0=0.001$\,\,\,\,\,Uncertainty of gluon at $x=0.001$ from fit to HERA+LHCb $J/\psi$ data} \par\medskip
\includegraphics[scale=0.65]{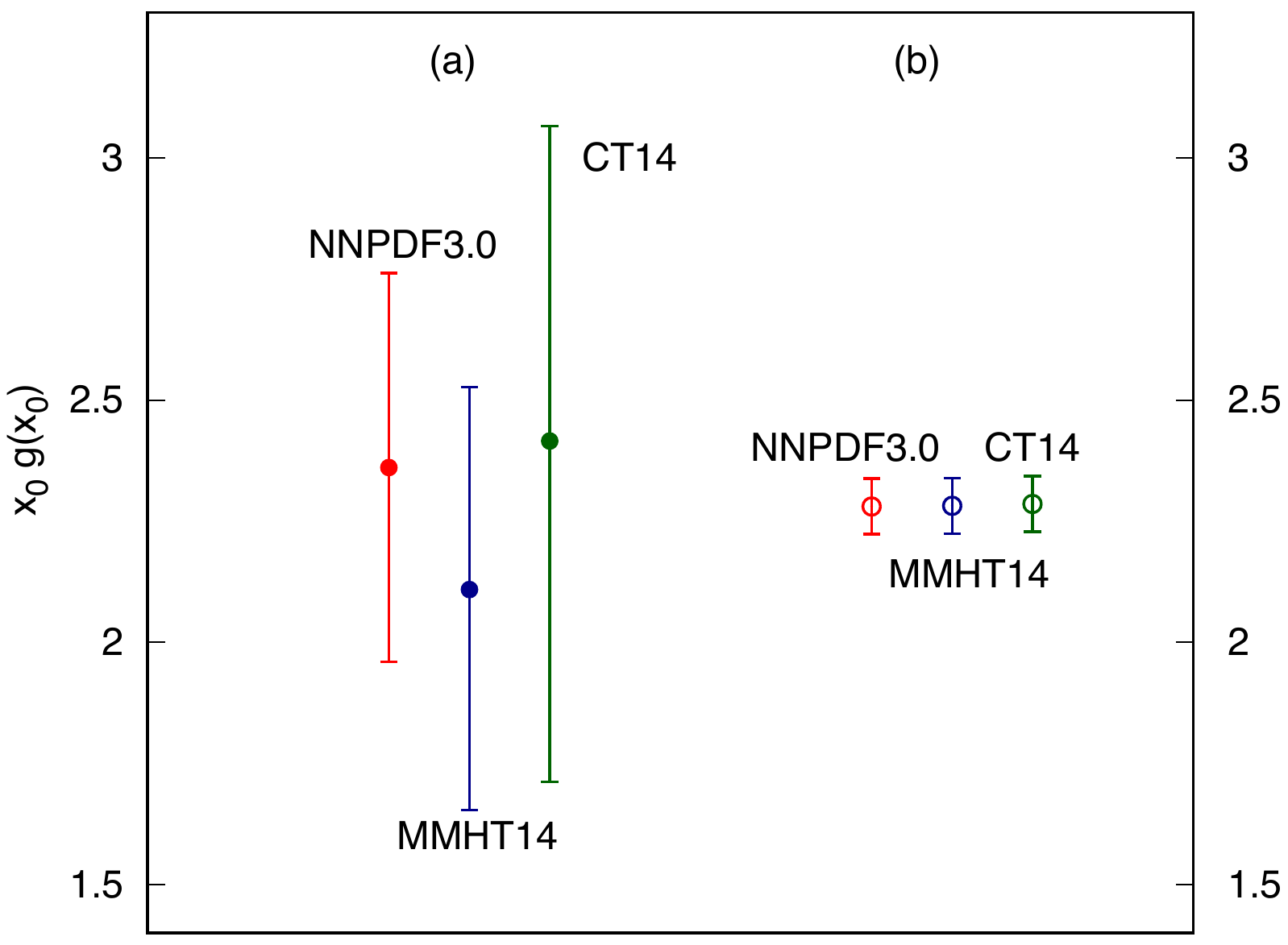}
%%\includegraphics[width=1.2\textwidth]{xg001.pdf}
%%\vspace{-6.0cm}
\caption{\sf{ (a) The global gluon PDF, $xg(x,\mu^2)$, at the matching point $x=0.001$ and $\mu^2=2.4$ GeV$^2$, (b) the global gluon PDF, $xg(x,\mu^2)$, at the matching point $x=0.001$ and $\mu^2=2.4$ GeV$^2$ after fitting to HERA+LHCb exclusive $J/\psi$ data. Note that the errors shown on the right hand side are those obtained by propagating the $1\sigma$ experimental data errors to our result, but do not account for theoretical uncertainties.  }}
\label{fig:xg001}
\end{center}
\end{figure}
%Possible new caption:  The gluon PDF, $xg(x,\mu^2)$, at the matching point $x=0.001$ and $\mu^2=2.4$ GeV$^2$ before and after the fit. ...
 
%So far we have based our analysis on matching to the central values of the global gluon PDFs at $x=0.001$.
%However these values have large uncertainties.
In the left hand side of Fig.~\ref{fig:xg001} we compare the uncertainties of the gluon densities given at $x=0.001$ and $\mu^2=2.4$ GeV$^2$ by the global analyses, while in the right hand side we show the values that are obtained after fitting the $J/\psi$ data.  The $J/\psi$ data are seen to greatly improve the knowledge of the gluon in the low $x$ interval $3\times 10^{-6}<x<10^{-3}$. In particular, we find at $x_0 = 0.001$ that
\be
x_0 g(x_0, \mu^2 = 2.4\,\text{GeV}^2) = 2.28 \pm 0.06.
%\,\,\,\,\,\,\,\,\,\,\,\,\,\,\,(\text{A})
\label{eq:central}
\ee

\subsection{The alternative double-log parametrization}

While the simple two parameter ansatz in~(\ref{eq:pow}) leads to a very good description of the $J/\psi$ data, it is still informative to repeat the procedure using the double-log ansatz in~(\ref{eq:kt}). Recall that a similar form was used in \cite{Jones}. The result obtained using the NNPDF3.0 NLO parton set is
%The description of the low $x$-$J/\psi$ data using the simple two-parameter ansatz of (4) is so good that no further parameters need to be introduced. 
%Moreover to obtain the low x behaviour of the gluon PDF, it is not necessary to consider other two-parameter forms.  
%However, it is informative to repeat the fit using the double-log ansatz given in (6) in which the parameters are $a$ and $n$.  (Recall that a similar form was used in [18].)  The result obtained using the NNPDF3.0 NLO parton set is
$$a = -0.046 \pm 0.006,\,\,\,\,\,n=0.979 \pm 0.025,\,\,\,\,\, \chi^2_{\text{min}}/\text{d.o.f} = 1.05.
$$
%The behaviour of the low $x$ gluon obtained for the two parametrizations, (4) and (6), are compared in Fig. 6.  
 The description and the behaviour of the low $x$ gluon are very similar to that obtained using~(\ref{eq:pow}).
 We find that the fit using the double log parametrization gives the central value $x_0 g(x_0, \mu^2=2.4 \, \text{GeV}^2) = 2.31$ in agreement with~(\ref{eq:central}).
 
 %that obtained using (\ref{eq:pow}).
 %but that the new fit, although acceptable, is worse in terms of $\chi^2_{\text{min}}.$ [NEED THIS SENTENCE?? - THE CHI SQ/D.O.F ARE REALLY MORE OR LESS THE SAME]

Note that the double-log parametrization gives a result close to that obtained in the $k_t$-factorization approach~\cite{Jones}.  However now, accounting for the complete set of NLO corrections, we find that the gluon growth with energy ($1/x$) is less steep than that obtained in~\cite{Jones}.
%a somewhat less steeper growth with energy ($1/x$) than that in [18]. 
Instead of $a= -0.10$ we now have $a \sim -0.05.$ The LHCb16 data 
%\cite{cc1} 
used in~\cite{Jones} have been replaced by the data in~\cite{LHCb} that is used here, but this is not accountable for the difference in $a$.

\subsection{Is there evidence of saturation from exclusive $J/\psi$ data?}
%{Discussion of the claim of evidence of saturation fromexclusive $J/\psi$ data

High energy exclusive $J/\psi$ production was recently described
in~\cite{1904} based on a BFKL approach. The authors claim that ``there
are strong hints for the presence of the saturation effects in exclusive
photo-production of
 $J/\psi$ at small $x$". We have to emphasize that actually the authors of~\cite{1904} refer to {\it absorptive
corrections} rather than  {\it saturation}. Indeed, saturation means
that the gluon density tends to a constant value, $xg(x,\mu^2)\to const$
as $x\to 0$ and at a fixed scale $\mu$~\cite{GLR}. That is, the power
$\lambda$ in (\ref{eq:pow}) behaves as
$\lambda\to 0$.
 A first hint of saturation would be to observe that the power
$\lambda$ (measured in some small-$x$ interval) starts to decrease with
decreasing $x$. The data, as shown in Fig. 3, do not indicate such behaviour. 
%On the contrary, the data are found to require a larger
%$\lambda=0.13-0.14$ at small $x<0.001$ than that ($\sim 0.06$) measured in
%the $0.001<x<0.01$ interval NEED TO CHANGE WORDING AND CORRECT.

 What is actually shown in~\cite{1904} is that the LO BFKL intercept,
$\alpha_{\text{BFKL}}=1+\omega_0=1+\lambda$
 is too large to describe the high energy $J/\psi$ data and  that
absorptive corrections (which are included into the non-linear
BK~\cite{BK} equation) are needed to tame the growth of the gluon density
(\ref{eq:pow}), that is to decrease the value of
$\lambda$.

 It is well known that the LO BFKL intercept is too large.
 It becomes smaller in the next-to-leading (NLL) approximation.
Indeed,  it is seen from \cite{1904} (the short dashed green curve of
their Fig.~1)  that the HSS gluons~\cite{HSS}, based on the NLO BFKL
{\em linear} equation, are
 in agreement with the exclusive $J/\psi$ data.

Therefore the growth of the gluon density with a smaller but non-zero
$\lambda$ is not evidence for `saturation'. At the moment no hint
of saturation is observed in exclusive $J/\psi$ data  at the scale
$\mu^2=2.4$ GeV$^2$ and $x$ down to 10$^{-5}$.

%From the point of view of future global fits, we find that the fit using the double-log parametrization gives
%$$
%x_0g(x_0, \mu^2 = 2.4\,\text{GeV}^2) = 2.13 \pm 0.04

%Besides the power like parametrization (\ref{eq:pow}) in Fig.\ref{fig:6} we show the result obtained using the replicas method constrained by $J/\psi$ data and based on the NNPDF3.0 global partons. \\

%{\bf Chris if we will show the results with Your 'J/psi-constrained' gluons then we have to describe how these gluons were obtained. What data ($\sigma$ or Ratios or both were used). Please, write the corresponding paragraph/subsection or Appendix}\\
%+++++++++++++++++++++

\subsection{Note on higher-twist contributions}

 Recall that absorptive corrections, which provide the saturation, are described by higher-twist operators. Formally, within the collinear factorization approach, we do not know the value of these higher-twist terms. 
 %The higher twist contributions 
 They have their own evolution and input conditions/functions that must be fitted from experiment. In other words, only experiment can give us the values of the higher-twist operator contributions.
Nevertheless, let us estimate the possible role of the higher-twist absorptive effects in the $J/\psi$ photoproduction amplitude.

The relative size of the contribution of the next twist absorptive correction (in our $\mu^2$ region of interest) is driven by the parameter (see~\cite{GLR})\footnote{In our approach everything below $Q_0$ (i.e. at scales $\mu^2< 2.4$ GeV$^2$) is
considered as a phenomenological input distribution which is formed
mainly by non-perturbative interactions inside the proton. We never go
below $Q_0$;  we subtract all the contributions with $k_t<Q_0$.  One therefore
cannot use our higher-twist estimate (of perturbative origin)  at lower scales.}
\be
c = \alpha_s \frac{x g(x)}{R^2 \mu_0^2},
\label{absorptive}
\ee
where $R$ can be as large as the proton radius $(R \sim 0.84~\text{fm})$. If we consider the value of $R$ as the `hot spot' radius\footnote{It may be assumed that the low-$x$ partons group together in so-called `hot-spots', with a radius smaller than that of the proton.}, then we have to take a smaller gluon density, $xg$, corresponding to only one hot spot. With $\alpha_s$ = 1/3 and $\mu_0 = M_{\psi}/2$ we obtain $c = 0.008\,xg \sim 0.04$ for our gluon density $xg \leq 5$.\footnote{A relatively large value of $xg = 5$ includes/accounts for the power growth of gluon densities at low $x$.}
However, actually this result is overestimated. Indeed, the cross section of an additional high energy (gluon) interaction is proportional to the $c$-quark
separation $\langle r^2 \rangle.$ This means that we have to replace in (\ref{absorptive}) the factor
$1/(R^2 \mu_0^2)$ by the ratio $\langle r^2 \rangle/R^2$. At the beginning of the photoproduction process, the photon produces a {\it point-like} $c\bar{c}$ pair. The lifetime of this pair is about $2E_{\gamma}/M_{\psi}^2$, where $E_{\gamma}$ is the energy of the photon. Accounting in addition for the Lorentz factor of the $J/\psi$, the quarks have their `own' time $\tau \sim 2/M_{\psi} = 1/\mu_0$ to separate from each other. However, the $J/\psi$ meson is a non-relativistic system and the heavy quark velocity $\langle v^2 \rangle \propto \alpha_s$ is small. That is we expect the higher-twist contribution to be suppressed by an additional power of $\alpha_s$ and, correspondingly, actually $c < 0.015.$ Accounting for the velocity $\langle v^2 \rangle \propto \alpha_s$ can be considered as a NNLO contribution.

\section{Comparison with low $x$ gluons from $D$-meson data}

As mentioned in the introduction, it is also possible to determine the low $x$ gluon
density from the data for various modes of inclusive open charm production of $D$-mesons and
their excited states. In this section, we provide a comparison of the results obtained from the data for inclusive $D$-meson
production and exclusive $J/\psi$ production. 
%is shown in Fig.~\ref{fig:6}.

\begin{figure} [t]
%%\vspace{-2.3cm}
\begin{center}
\includegraphics[scale=0.7]{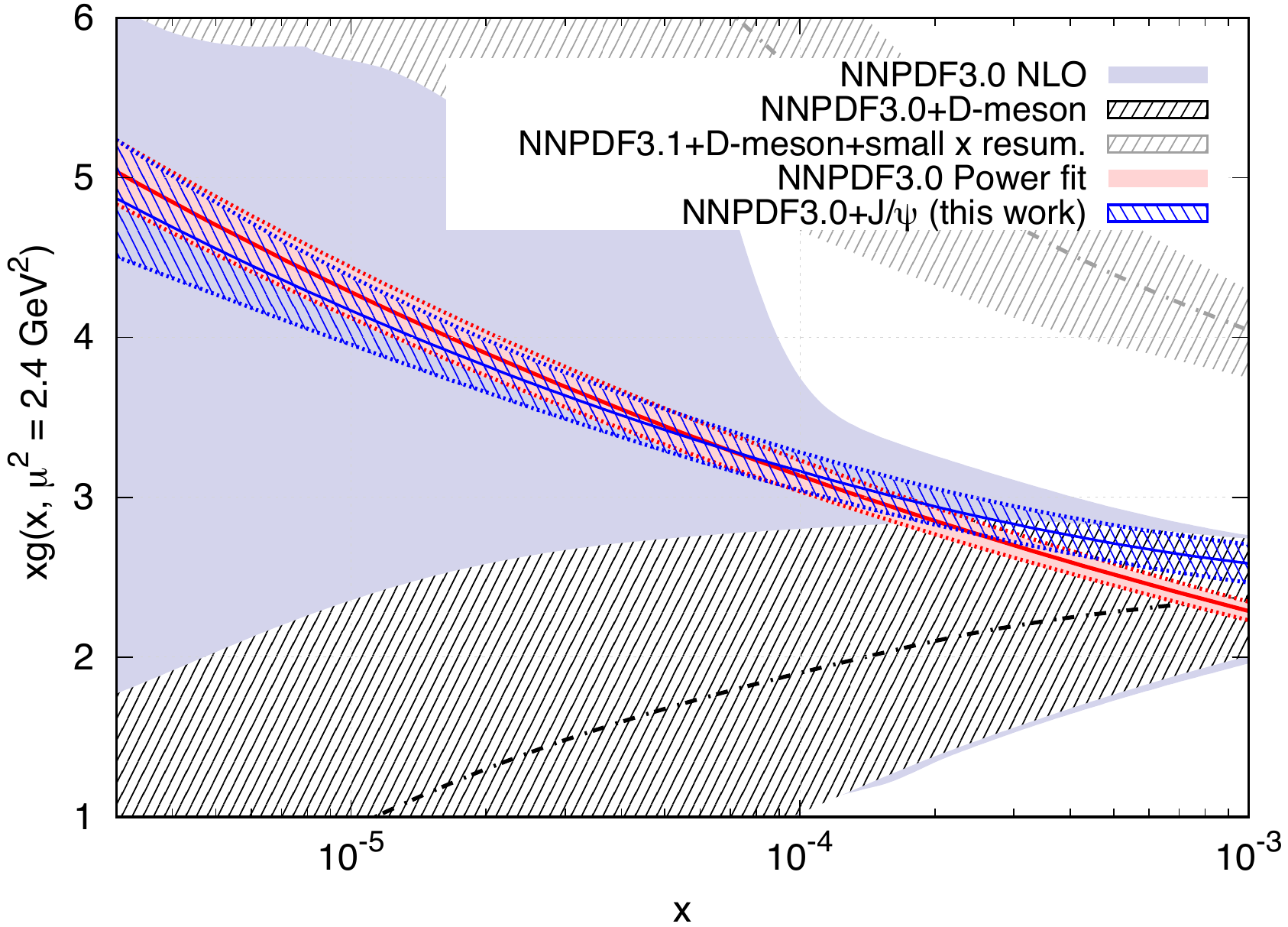}
%%\includegraphics[width=0.75\textwidth]{Fig6.pdf}
%\includegraphics[width=0.75\textwidth]{Fig6a.pdf}
%\includegraphics[width=0.49\textwidth]{chisqDL070220.pdf}
%\includegraphics[width=0.4\textwidth]{Fig2Quark.pdf}
%\vspace{-5.0cm}
\caption{\sf{Comparison of the low $x$ behaviour of the NLO gluon density $xg(x,\mu^2)$ at $\mu^2=2.4$ GeV$^2$ obtained from exclusive $J/\psi$ data and from inclusive $D$-meson data, see text for details.}}
%{\bf I propose to use the first figure but with the error band for power gluons corresponding to +/-10\% from the 2d figure. Do you agree ?}}}
\label{fig:6}
\end{center}
\end{figure}

Inclusive $D$-meson production data via $pp$ collisions at the LHC are available at
centre of mass energies $5,7$ and $13\; \text{TeV}$~\cite{cc1}. The kinematics of the different
modes of production of the $D$-mesons allow for a coverage down to $x\sim{\rm few}\times 10^{-6}$.
In \cite{r7}
%[Gauld,Rojo],   
the authors studied the impact these data for $\left\{D^0, D^+, D^+_s\right\}$ final states would have on the small
$x$ NLO gluon within the NNPDF3.0 global analysis through a Bayesian reweighting. While
the corresponding NLO calculation for $D$-meson production suffers from large theory
uncertainties attributed to the dependence on the factorization scale and large higher order corrections, construction of
ratios of the double-differential cross section in rapidity and transverse momentum
bins provides a means to combat this residual scale dependence and thereby
quantitatively assess the impact the data would have in the PDF fit. Of course, the
overall normalisation is forfeited but the sensitivity to the $x$ dependence of the gluon
is maintained in this approach. In Fig.~\ref{fig:6} we show the NNPDF3.0
global gluon reweighted using the ratios of inclusive $D$-meson cross section data at $\sqrt{s} = 5,7,13~\text{TeV}$ and 
%for the baseline data combination $N_{5+7+13}$ 
evolved down to the $J/\psi$ scale $\mu^2 = 2.4
\,\text{GeV}^2$ (the lower grey band).  As shown and explained
 in~\cite{r7}, % [Gauld,Rojo], 
 the data favour a
decreasing gluon at the lowest value of $x$ which the $D$-meson data may probe.

\begin{figure} [t]
%%\vspace{-2.3cm}
\begin{center}
\includegraphics[scale=0.7]{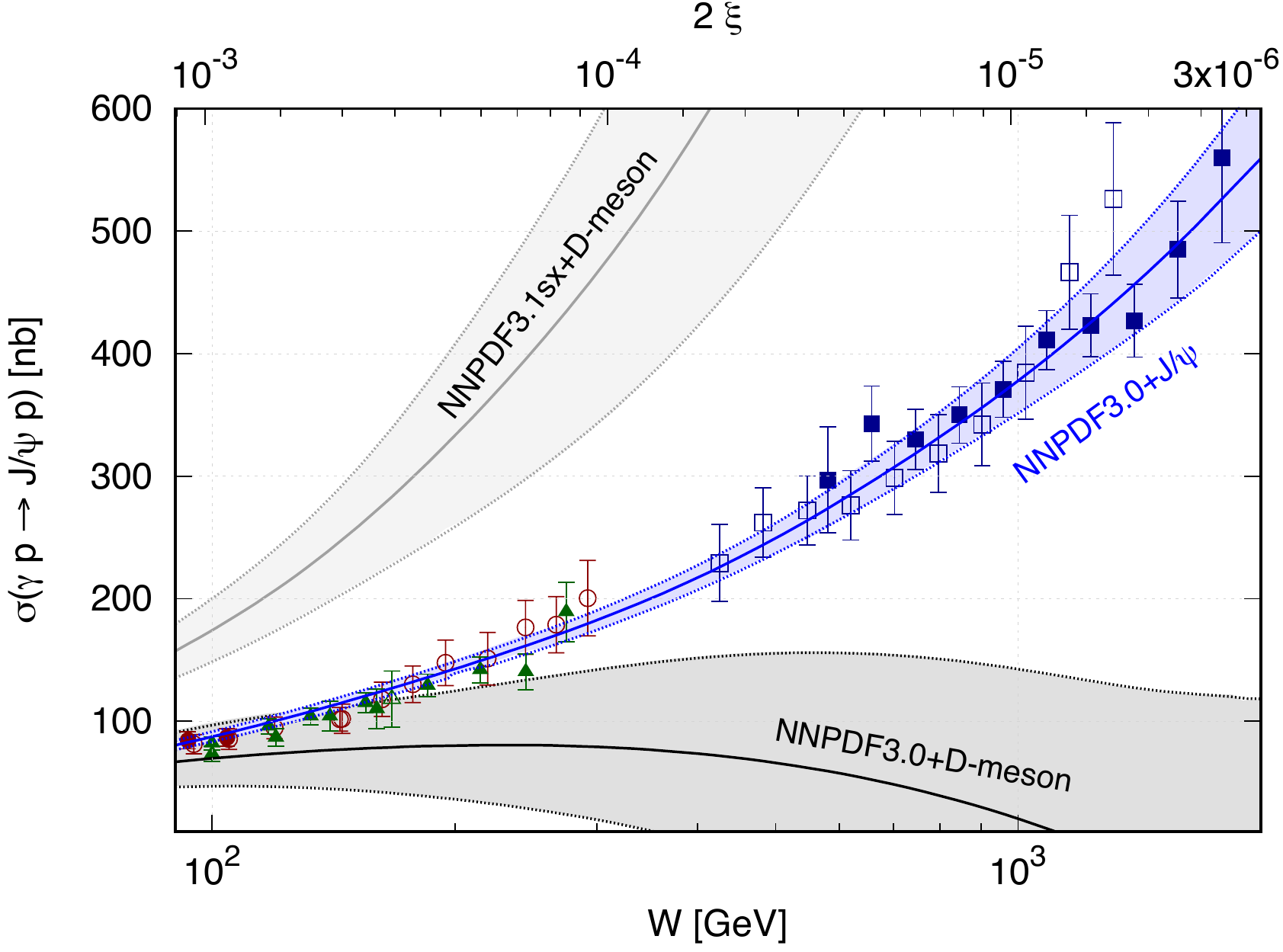}
%%\includegraphics[width=0.75\textwidth]{Fig6.pdf}
%\includegraphics[width=0.75\textwidth]{Fig6a.pdf}
%\includegraphics[width=0.49\textwidth]{chisqDL070220.pdf}
%\includegraphics[width=0.4\textwidth]{Fig2Quark.pdf}
%\vspace{-5.0cm}
\caption{\sf{The lower and upper bands are, respectively, the cross section predictions obtained using NNPDF3.0 and NNPDF3.1 global partons constrained by the $D$-meson LHCb data \cite{ r7, Bertone}. The latter includes low $x$ resummation effects. The shaded blue band is the cross section prediction obtained based on our reweighting of the NNPDF3.0 NLO global gluon via the exclusive $J/\psi$ data. The experimental data points are presented as in Fig.~\ref{fig:2}. }}
%{\bf I propose to use the first figure but with the error band for power gluons corresponding to +/-10\% from the 2d figure. Do you agree ?}}}
\label{fig:7}
\end{center}
\end{figure}

This is to be contrasted with the same analysis performed for NNPDF3.1 supplemented
with the inclusive $D$-meson data but now together with small $x$ resummation~\cite{Bertone}.
%\footnote{The authors would like to thank Valerio Bertone for private communications and for providing us with these constraints in LHAPDF6 format <---OR LET US PUT IN ACKNOWLEDGEMENTS.}. % ref
%[BertoneEtAl]. 
In this case, the reweighting favours a much higher gluon, as shown
by the upper grey band in Fig.~\ref{fig:6}. It is known that including the BFKL (small $x$) resummation ({\em without a $k_t<Q_0$ subtraction}) the low scale gluons extrapolated into the low $x<0.001$ region are too large and grow too fast (see e.g.~\cite{LLx}). That is, as shown in Fig.~\ref{fig:7},  the cross section prediction using NNPDF3.1 together with the resummation strongly overshoots the exclusive $J/\psi$ data while the prediction using NNPDF3.0 
%the obtained $\sigma(\gamma p\to J/\psi p)$ 
is too low. 

The comparison of these two (based on NNPDF3.0 and on NNPDF3.1) bands, together with the inconsistencies of $D$-meson data mentioned in~\cite{Gauld,OMR}, demonstrates that the quality and accuracy of $D$-meson data are not sufficient to get an unambiguous result and to obtain accurate low $x$ gluons. 
%Recall also the inconsistencies of $D$-meson data mentioned in~\cite{Gauld,OMR}.

\section{Discussion}

In this work, we too have performed a Bayesian reweighting of the NNPDF3.0 gluon but
this time constrained by the exclusive $J/\psi$ cross section. As discussed in~\cite{PRD} 
% [ref our previous paper], 
these data are in a position to be readily included in a collinear NLO global
analysis due to alleviation of the large scale dependence through implementation of
a $Q_0$ cut and resummation of a class of large logarithms. We %systematically
have performed the reweighting using the $J/\psi$ data in the region $x < 0.01$ for the NNPDF3.0 NLO set with $N_{\text{rep}} = 1000$ replicas.
%reweighting using the NNPDF3.0 NLO set with $N_{\text{rep}} = 1000$ replicas.
%1000 %(1000?) replica set of NNPDF3.0 NLO. 
Since the central NNPDF3.0 low $x$ gluons are too large to describe the $J/\psi$ data (see Fig.~\ref{fig:2}), the Shannon entropy (or effective number of contributing replicas), $N_{\text{eff}} \approx 40 \ll N_{\text{rep}}$.
%, where $N_{\text{rep}}$ is the number of NNPDF3.0 replicas. 
Therefore, the reweighting approach is not fully adequate. Still, the obtained gluons (hatched blue band in Fig.~\ref{fig:6}) are rather close to that obtained within the fit using ansatz~(\ref{eq:pow}). 
%{\it Since the form of the input gluon distribution used in the global NNPDF3.0 fit does not allow for different powers of $\lambda$ at very small $x<0.001$ and inside the $0.001<x<0.01$ interval, the $J/\psi$ reweighted NNPDF3.0 gluons have a bit smaller value of $\lambda$ in comparison with that coming from the power fit (\ref{eq:pow}). Correspondingly, the $J/\psi$ reweighted gluon density overshoots our (power fit) result at $x=x_0=0.001$ and undershoots it at the smallest $x=3 \times 10^{-6}$} - NEED TO DELETE/CORRECT?. 
Since the NNPDF input distribution is mainly driven by other data at larger $x\sim 0.01$ (where
the effective value of $\lambda$ is noticeably smaller), the reweighted NNPDF3.0 gluon has a slightly less steep growth at $x<0.001$ in comparison with that coming from the power fit~(\ref{eq:pow}). Correspondingly, the $J/\psi$ reweighted gluon density overshoots our (power fit) result at $x=x_0=0.001$ while undershooting it at the smallest $x=3 \times 10^{-6}$.\footnote{The slightly larger normalization, at $x=10^{-3}$, of the prediction based on the reweighting procedure is due to the greater number of data points that are fitted in this region in the global analysis. For smaller $x$, where the only constraining power comes from the exclusive $J/\psi$ data in both the reweighting and power fit approaches, the predictions are in better agreement.} 
On the other hand our $J/\psi$ reweighting result demonstrates %, in our case,
 that 
the additional $J/\psi$ data adds a lot of new information, which is to be expected as there
were no data in the previous PDF analyses in this domain. The small
value of the Shannon entropy means it would be desirable for the
reweighting procedure to be backed up by a full new global fit. This quantifies the
statements in~\cite{PRD} % [ref our previous paper]
 about the utility of the $J/\psi$ data. The
closeness of our reweighted gluon with the fitted gluons we have obtained provides further support for this claim. Considering all data points with $W > 100 \,\text{GeV},$ the effective $\chi^2_{\text{min}}/N_{\text{dat}} \sim 1.07$ for the reweighted central cross section prediction.
%a slight inflation of that found for the NNPDF3.0 power fit due to the mismatch of the gluons obtained at $x=10^{-3}$. 
%[MAY ADD AN EFFECTIVE CHI SQ VALUE FOR REWEIGHTING]

Thus exploiting the $J/\psi$ exclusive data
 we reach a much better accuracy. Now, down to $x=3 \times 10^{-6}$, the low scale gluons
 (near the input $Q_0$ value) are known to better than 5-7\% uncertainty. 

 An interesting observation is that in the low $x<0.001$ region, the low scale fitted gluons start to grow (with $1/x$) even faster (as $xg(x)\propto x^{-\lambda}$ with $\lambda \simeq 0.14$) than the low scale global gluons do
 %it was before 
 in the interval $0.001<x<0.01$. We are able to fit a low $x$ gluon power ansatz for the large range $x<0.001$ with a {\it single} slope but find that we cannot extend this same description to $0.001<x<0.01$. Attempting to do so results in a worsened fit and a much smaller $\lambda$. 
 %(in attempting to do so results in a different power behaviour and much higher $\chi_{\text{min}}^2/\text{d.o.f}$).
 Indeed, this reflects the differing behaviour
 %effective lambda
 of the NLO global gluons in the intervals $0.001<x<0.01$ and $x<0.001$.  The fact that the effective power $\lambda$ {\em increases} with $1/x$ (within the $10^{-2} - 10^{-5}$ interval) is in contradiction with the assumption of saturation for which one would expect a decreasing $\lambda \to 0$ as $x \to 0$.
%saturation regime where one expects the decreasing $\lambda\to 0$ as $x\to 0$.
 %That is to say that the 
 %Indeed, fitting the data for $W > 30$ GeV with $x_0=0.01$ using the power ansatz (5) we obtain a much smaller $\lambda \sim 0.06$ and much higher $\chi_{\text{min}}^2/\text{d.o.f}=2.9$. That is to say 
The data with $x<0.01$, therefore, cannot be described by a single power behaviour, indicative of non-trivial non-perturbative effects in the input proton wave function.
 %in the region $0.001<x<0.01.$ 
 %[NEED TO SAY MORE OR CHANGE/CORRECT??].
 
 %{\bf Chris, can you, please, to check and to insert the numbers insteed of "???". Thank you.}}. 
 On the other hand note that the power $\lambda \simeq 0.14$ (that we obtained in the description of the $J/\psi$ data with $x<0.001$) is close to that predicted by the NLL BFKL re-summed  with the optimal (BLM~\cite{BLM}) scale renormalization~\cite{Kim}. Moreover, contrary to the common expectation, even at $x\sim 10^{-5}$ and $\mu^2=2.4$ GeV$^2$, in our approach we see no hint in the exclusive $J/\psi$ data for the onset of parton density saturation.

\section{Conclusion}
High energy HERA and LHCb data on exclusive $J/\psi$ production were described using a consistent collinear factorization approach at NLO. We fix the `optimal' factorization scale $\mu_F=M_{\psi}/2$, which allows for the resummation of the double-logarithmic $(\alpha_s\ln(1/x)\ln\mu_F)^m$ corrections into the incoming PDF, and subtract the low $k_t<Q_0$ contribution from the coefficient function to avoid double counting between the NLO coefficient function and the contribution hidden in the input PDF (or GPD) at $Q=Q_0$. This provides good stability of the results with respect to variations of $\mu_f$.  The generalized GPD distribution was related to the conventional (non-skewed) PDF via the Shuvaev transform. The renormalization scale is $\mu_R=\mu_f$.

With this, we find collinear NLO gluons at $\mu^2=2.4$ GeV$^2$ %(very close to the input $Q_0$ value) 
which give an excellent description of all available accurate $J/\psi$ data throughout the very low $x$ interval, $3\times 10^{-6} < x< 10^{-3}$, to about $\pm$~5-7~\% accuracy at the lowest $x$. 
 The gluon PDF $xg(x,\mu^2)\propto x^{-\lambda}$ increases with $1/x$ with $\lambda=0.135\pm 0.006$ without any hint in favour of parton density saturation at $\mu^2 = 2.4\,\text{GeV}^2$ and $x$ down to $10^{-5}$. We emphasize this does not
mean that the data cannot be described by a more complicated
expression which ultimately (at very small $x$) will provide
saturation.
 
 A Bayesian reweighting approach leads to a similar behaviour of the small $x$ gluon, emphasising the utility and constraining power of the exclusive $J/\psi$ data. This work therefore clearly demonstrates the gains which will be achieved once these data are included in the global PDF fits.

\section*{Acknowledgements}
We thank Stephen P. Jones for his major contributions (both theoretical and phenomenological) to our previous analyses of exclusive $J/\psi$ production which form much of the basis for the present study, and for his careful reading of our manuscript. The authors would also like to thank Valerio Bertone for a useful discussion and for providing the $D-$meson constrained NNPDF parton sets in \texttt{LHAPDF6} format. 
%We thank Stephen P. Jones for his major contributions (both theoretical and phenomenological) to our previous analyses of exclusive $J/\psi$ production which form much of the basis for the present study, and for his careful reading of our manuscript.
%We also thank Stephen P. Jones for useful discussions and for a critical reading of the manuscript. 
C.A.F and M.G.R thank the IPPP at Durham University for hospitality. The work of C.A.F is supported by an STFC award grant ST/N504130/1 and that of T.T is supported by STFC under the consolidated grants ST/P000290/1 and ST/S000879/1.

%JETP Letters,  S.J. Brodsky, V.S. Fadin, V.T. Kim, L.N. Lipatov, G.B. Pivovarov, {\bf 70} (19991) 155.

\end{document}